# Shared Lateral Control with On-Line Adaptation of the Automation Degree for Driver Steering Assist System: A Weighting Design Approach*

Anh-Tu Nguyen, Chouki Sentouh, Jean-Christophe Popieul and Boussaad Soualmi

*Abstract*— This paper addresses the shared lateral control for both lane-keeping and obstacle avoidance tasks of a driver steering assist system (DSAS). In this work, we propose a novel approach to deal with the interactions between the human (driver) and the machine (DSAS) by introducing into the vehicle system a fictive nonlinear term representing the driver activity. In this way, the actions of the DSAS are computed according to the driver behaviors (actions and intentions). Based on Takagi-Sugeno control technique together with Lyapunov stability tools, the designed controller is able to handle a large range of variation of vehicle longitudinal speed. In particular, this controller can deal with the system state constraints and also the control input saturation. As will be discussed later, the consideration of these constraints into the control design improves significantly the closed-loop performance under various driving situations. The interests of the proposed method are validated by simulations.

I. INTRODUCTION

Most of modern vehicles are equipped with a DSAS which can assist the driver in different driving situations. It has been proved that these systems contribute to reduce the workload of the driver and also the road accidents [1]. As a consequent, the control design of intelligent vehicle systems have attracted a great deal of attention from both academic and industrial researchers [2]–[5].

From the viewpoint of Human-Machine interactions, the introduction of a DSAS involves crucial impacts on driving process. Indeed, the DSAS acts on the steering system to modify the driver activity and also eventual risks coming from external environment. To this end, in many driving situations the conflict issues, i.e. when the driver and the DSAS do not have the same objectives, may arise. To prevent these situations, cooperative driving control strategies considering directly the Human-Machine interactions seem to be a promising solution [6]. The objective of such a control strategy is that the designed actions of the E-copilot do not prevent the driver to perform some specific tasks such as obstacle avoidance or lane-change maneuvers that have not been detected by the E-copilot or, even better, the E-copilot should help the driver to realize these maneuvers. Up to now, several studies have pointed out the need for an active coordination of authority between human driver and driving assistance systems [6]–[8]. In addition, existing works [6], [7], [9] have also shown that the integration of the driver behaviors in the control loop helps to improve effectively the conflict management between these two agents.

In our previous work [6], an interesting approach for vehicle lateral control that can share responsibility with driver was proposed to deal with both lane-keeping and obstacle avoidance tasks. Using the control technique based on Takagi-Sugeno (T-S) models [11], this approach is able to take into account a large range of variation of vehicle longitudinal speed. This consideration is crucial to guarantee a good closed-loop performance of the designed controller and also to allow for a better Human-Machine coordination under different driving situations. In that work, the cooperation degree is quantified by a linear quadratic (LQ) criterion to be minimized. However, the designed actions of the E-copilot cannot exploit the information on the driver behaviors and also his/her driving state coming from the supervision level. In this paper, we take a step further to design an "intelligent" controller for the DSAS which can overcome these drawbacks. To do that, a weighting factor (viewed as a system time-varying parameter) representing the driver activity will be judiciously introduced in the road-vehicle model. Considering directly this information in the control design seems very promising to solve the challenging conflict issue in the framework of shared lateral control. In order to deal with the variation of vehicle longitudinal speed and the introduced time-varying parameter, the control technique based on T-S model is employed. In particular, this control approach can guarantee various closed-loop properties in presence of actuator saturation, limitations on system states and also persistent disturbances bounded in amplitude. All design conditions are expressed in terms of linear matrix inequalities (LMIs) which can be effectively solved with numerical solvers [14].

For the proposed approach, it should be stressed that the control input saturation is not directly related to the physical actuator limitations of the steering system. This fact is due to the multiplication of the designed fictive torque and the weighting factor representing the driver activity. The bounds of this fictive torque are determined such that the whole capacity of the steering actuator (under the impacts of the weighting factor) can be exploited for the control performance. In many specific driving situations, such as

* This work has been done in the context of the CoCoVeA research program (ANR-13-TDMO-0005), funded by the National Research Agency. This work was also sponsored by the International Campus on Safety and Intermodality in Transportation, the Nord-Pas-de-Calais Region, European Community, the Regional Delegation for Research and Technology, the Ministry of Higher Education and Research, and the French National Center for Scientific Research.

AnhTu Nguyen, Chouki Sentouh, Jean-Christophe Popieul are with the laboratory LAMIH-UMR CNRS 8201, University of Valenciennes, France. Boussaad Soualmi is with the Institute for Technological Research SystemX, France. E-mail addresses: nguyen.trananhtu@gmail.com, {chouki.sentouh, jean-christophe.popieul}@univ-valenciennes.fr and boussaad.soualmi@irt-systemx.fr.

lane-change or obstacle avoidance maneuver (in such a situation it is assumed that the driver must be fully conscious of his/her driving actions), the fictive torque tends easily to be saturated to compensate the important torque received from the driver. Therefore, considering the input saturation in the control design is therefore crucial for this application to prevent the loss of closed-loop stability [12], [13]. To the best of our knowledge, these results were not observed in any previous works.

## II. MODELING FOR SHARED LATERAL CONTROL

Hereafter, only some highlights related to the control-based model are reviewed. The vehicle parameters used in this work are given in Table 1.

Table 1. Vehicle parameters

| Parameter | Description | Value |
|---|---|---|
| $M$ | Total mass of the vehicle | 1500 kg |
| $l_f$ | Distance from the GC to front axle | 1.0065 m |
| $l_r$ | Distance from the GC to rear axle | 1.4625 m |
| $l_w$ | Distance from the GC to impact center of the wind force | 0.4 m |
| $l_s$ | Look-ahead distance | 5 m |
| $\eta_t$ | Tire length contact | 0.13 m |
| $I_z$ | Vehicle yaw moment of inertia | 2454 kgm$^2$ |
| $I_s$ | Steering system moment of inertia | 0.05 kgm$^2$ |
| $R_s$ | Steering gear ratio | 16 |
| $B_s$ | Steering system damping coefficient | 15 |
| $C_f$ | Front cornering stiffness | 94270 N/rad |
| $C_r$ | Rear cornering stiffness | 113272 N/rad |
| $C_x$ | Longitudinal aerodynamic drag coefficient | 0.35 |
| $C_y$ | Lateral aerodynamic drag coefficient | 0.45 |
| $\tau_a$ | Time to tangent point | 0.5 s |
| $T_p$ | Driver preview time | 0.8 s |

### A. Simplified Road-Vehicle Model with Steering System

Based on the well-known "bicycle model" [1], the road-vehicle model integrating the electrical steering assistance system can be described by [6]:

$$\dot{x}_v = Ax_v + B(T_c + T_d) + B_w w \quad (1)$$

where the state vector $x_v^T = \begin{bmatrix} v_y & r & \psi_l & y_l & \delta & \dot{\delta} \end{bmatrix}$ is composed by the lateral velocity $(v_y)$, the yaw rate $(r)$, the heading error $(\psi_l)$, the lateral offset from the road centerline at a look-ahead distance $l_s$ $(y_l)$, the steering angle $(\delta)$ and its time derivative $(\dot{\delta})$. For system (1), the E-copilot torque $T_c$ has to be designed whereas the driver torque $T_d$ is measured. The lateral wind force $f_w$ is considered as disturbance. The system matrices are given by:

$$A = \begin{bmatrix} a_{11} & a_{12} & 0 & 0 & b_1 & 0 \\ a_{21} & a_{22} & 0 & 0 & b_2 & 0 \\ 0 & 1 & 0 & 0 & 0 & 0 \\ 1 & l_s & v_x & 0 & 0 & 0 \\ 0 & 0 & 0 & 0 & 0 & 1 \\ T_{s1} & T_{s2} & 0 & 0 & T_{s3} & T_{s4} \end{bmatrix}, B = \begin{bmatrix} 0 \\ 0 \\ 0 \\ 0 \\ 0 \\ \rho \end{bmatrix}, B_w = \begin{bmatrix} e_1 \\ e_2 \\ 0 \\ 0 \\ 0 \\ 0 \end{bmatrix}, \quad (2)$$

where

$$a_{11} = -\frac{C_r + C_f}{Mv_x}, \, a_{12} = -v_x + \frac{l_r C_r - l_f C_f}{Mv_x}, \, b_1 = \frac{C_f}{M}, \, e_1 = \frac{1}{M},$$

$$a_{21} = \frac{l_r C_r - l_f C_f}{I_z v_x}, \, a_{22} = -\frac{l_r^2 C_r + l_f^2 C_f}{I_z v_x}, \, b_2 = \frac{l_f C_f}{I_z}, \, e_2 = \frac{l_w}{I_z} \quad (3)$$

$$T_{s1} = \frac{C_f \eta_t}{I_s R_s^2 v_x}, \, T_{s2} = \frac{C_f l_f \eta_t}{I_s R_s^2 v_x}, \, T_{s3} = -\frac{C_f \eta_t}{I_s R_s^2}, \, T_{s4} = \frac{-B_s}{I_s}, \, \rho = \frac{1}{I_s R_s}.$$

### B. Driver-in-the-Loop Vehicle Model

It has been shown in [6], [7] that the integration of the driver behaviors into the control design procedure plays a key role for shared lateral control in terms of conflict management between the E-copilot and the driver. For lane keeping and/or obstacle avoidance tasks, a simplified driver model [7] is used. To this end, the driver torque can be modeled as a linear combination of the near viewpoint angle $\theta_{near}$ and the angle representing the direction of the car heading and the tangent point $\theta_{far}$:

$$T_d = K_{d1} \theta_{near} + K_{d2} \theta_{far}, \quad (4)$$

where the two gains $K_{d1}$, $K_{d2}$ characterizing the driver style are identified with real-time data. The expressions of two visual angles are given as:

$$\begin{cases} \theta_{near} = \frac{y_l}{v_x T_p} + \psi_l, \quad \theta_{far} = \theta_1 v_y + \theta_2 r + \theta_3 \delta_d, \\ \theta_1 = \tau_a^2 a_{21}, \quad \theta_2 = \tau_a + \tau_a^2 a_{22}, \quad \theta_3 = \tau_a^2 b_2, \quad \delta_d = \delta R_s. \end{cases} \quad (5)$$

From (4)-(5), it follows that:

$$T_d = T_{d1} v_y + T_{d2} r + K_{d1} \psi_l + T_{d3} y_l + T_{d4} \delta, \quad (6)$$

where $T_{d1} = K_{d2} \tau_a^2 a_{21}$, $T_{d2} = K_{d2} (\tau_a + \tau_a^2 a_{22})$, $T_{d3} = K_{d1}/(v_x T_p)$ and $T_{d4} = K_{d2} \tau_a^2 b_2 R_s$. By integrating the driver model (6) into road-vehicle system (1), the diver-in-the-loop model used for shared lateral control can be represented as:

$$\dot{x}_v = A_v x_v + BT_c + B_w w \quad (7)$$

where

$$A_v = \begin{bmatrix} a_{11} & a_{12} & 0 & 0 & b_1 & 0 \\ a_{21} & a_{22} & 0 & 0 & b_2 & 0 \\ 0 & 1 & 0 & 0 & 0 & 0 \\ 1 & l_s & v_x & 0 & 0 & 0 \\ 0 & 0 & 0 & 0 & 0 & 1 \\ T_{s1} + \rho T_{d1} & T_{s2} + \rho T_{d2} & K_{d1} \rho & \rho T_{d3} & T_{s3} + \rho T_{d4} & T_{s4} \end{bmatrix}.$$

## III. INTELLIGENT COOPERATIVE CONTROL STRATEGY

This work aims at proposing an "intelligent" sharing control strategy for the DSAS to assist the driver in different driving situations. To this end, the resulting strategy must be able to take into account the interactions between the driver and the E-copilot to manage effectively all their eventual conflicts. In [8], the authors have provided an interesting study on the need for assistance according to the driver load and driver performance, see Figure 1. It can be observed that the assistance levels should be designed to relieve the driver in overload and underload conditions. The cooperative control strategy proposed in this work follows this idea.

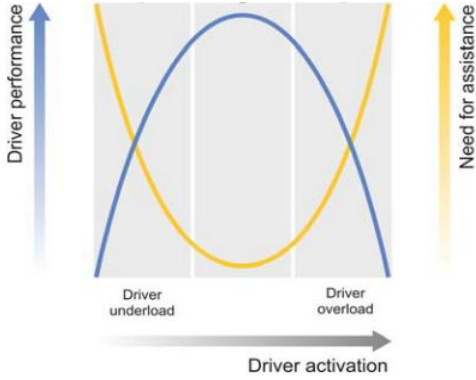

Figure 1. U-shape function representing the need for assistance according to driver load and performance [8]

In order to realize the above cooperative control strategy, we propose to modulate the assistance torque $T_c$ in function of the driver activity as follows:

$$T_c = \mu(\theta_d)u, \quad (8)$$

where the fictive torque $u$ is computed by the controller and the normalized variable $\theta_d$ represents the driver activity as discussed later. Notice that the weighting function $\mu(\theta_d)$ in (8) allows for a continuous assistance from the DSAS. In this work, the weighting factor $\mu(\theta_d)$ is inspired from the generalized Bell-shape function of the form:

$$\mu(\theta_d) = \frac{1}{1 + \left|\frac{\theta_d - \omega_3}{\omega_1}\right|^{2\omega_2}} + \mu_{\min} \quad (9)$$

where the parameters $\omega_1$, $\omega_2$, $\omega_3$ and the minimal assistance level $\mu_{\min}$ are parameterized to "mimic" the U-shape function, see Figure 2. From (7) and (8), the driver-in-the-loop vehicle model can be rewritten as:

$$\begin{cases} \dot{x}_v = A_v x_v + B_u u + B_w w \\ y_v = C x_v \end{cases} \quad (10)$$

where $B_u^T = \begin{bmatrix} 0 & 0 & 0 & 0 & 0 & \mu(\theta_d)\rho \end{bmatrix}$. Note that since $\mu(\theta_d) > 0$, $\forall \theta_d$, the controllability of system (10) is always guaranteed.

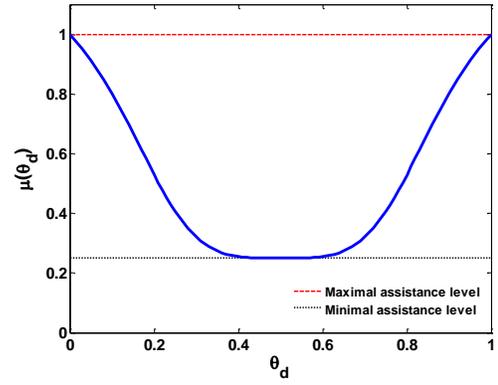

Figure 2. Weighting factor as a function of driver activity with $\omega_1 = 0.38$, $\omega_2 = -2$, $\omega_3 = 0.5$ and $\mu_{\min} = 0.25$

For the proposed method, the driver activity variable $\theta_d$ depends on two factors: the measured driver torque and the driver state information coming from driver monitoring unit ($0 \leq \text{DS} \leq 1$). This normalized variable $\theta_d$ is of the form:

$$\theta_d = 1 - e^{-(\sigma_1 T_{d,N})^{\sigma_2} \cdot \text{DS}^{\sigma_3}} \quad (11)$$

The normalized driver torque is defined as $T_{d,N} = |T_d/T_{d,\max}|$ where $T_{d,\max}$ is the maximal torque can be delivered by the driver. The tuning parameter $\sigma_1$ is used together with $T_{d,N}$ to represent the involvement level of the driver in the driving tasks whereas the parameters $\sigma_2$ and $\sigma_3$ represent the degree of influence of the driver torque and the driver state on the driver activity variable $\theta_d$. As can be observed in Figure 3, when the driver torque and/or the driver state remain small (which means that driver activity is not significant), the corresponding values of the normalized variable $\theta_d$ are small and high level of assistance is required. When the driver is highly involved in his/her driving tasks, i.e. $\theta_d$ tends to 1 (for example in the case where the driver need to realize a difficult driving task), important assistance from the DSAS is also required to help him/her. With a constant driver torque (respectively driver state), the driver activity increases according to the driver state (respectively driver torque).

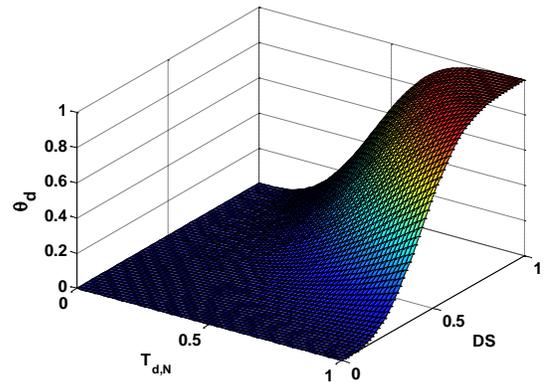

Figure 3. Driver activity function with $\sigma_1 = 2$, $\sigma_2 = \sigma_3 = 3$

## IV. TAKAGI-SUGENO MODEL-BASED CONTROL DESIGN FOR COOPERATIVE DRIVING TASKS

In this section, the design conditions for input-saturated T-S systems are derived by using Lyapunov stability tools. Then, application to the shared lateral control issue is given.

### A. Control Design for Input-Saturated T-S Systems

Consider the following Takagi-Sugeno model [11]:

$$\begin{cases} \dot{x} = \sum_{i=1}^{r} \eta_i(\theta)\left(A_i x + B_i^u \text{sat}(u) + B_i^w w\right) \\ z = \sum_{i=1}^{r} \eta_i(\theta) C_i x \end{cases} \quad (12)$$

where $x \in \mathbb{R}^{n_x}$, $u \in \mathbb{R}^{n_u}$, $w \in \mathbb{R}^{n_w}$ and $z \in \mathbb{R}^{n_z}$ are respectively the state, the control input, the disturbance and the performance output vectors of the system. The scheduling variable vector $\theta \in \mathbb{R}^k$ is known and it may be a function of the system states. The saturation function $\text{sat}(u)$ is defined by $\text{sat}(u_{(i)}) = \text{sign}(u_{(i)}) \min(|u_{(i)}|, u_{\max(i)})$ where $u_{\max(i)}$, $i \in \{1,\ldots,n_u\}$, denotes the amplitude bound relative to the $i^{th}$ control input. The real constant matrices of appropriate dimensions $A_i$, $B_i^u$, $B_i^w$, $C_i$, $i \in \{1,\ldots,r\}$, represent the set of $r$ local linear subsystems and the scalar membership functions $\eta_i(\theta)$ satisfy the property:

$$\sum_{i=1}^{r} \eta_i(\theta) = 1, \quad \eta_i(\theta) \geq 0, \quad \forall i \in \{1,\ldots,r\}. \quad (13)$$

For system (12), it is assumed that the disturbance signal $w$ is bounded in amplitude:

$$\mathcal{W}_\rho = \{w : \mathbb{R}^+ \mapsto \mathbb{R}^{n_w}, \ w^T R w \leq \rho\} \quad (14)$$

where the matrix $R > 0$ and the bound $\rho > 0$ are given.

Let us consider the parameter-dependent control law of the form:

$$u = \sum_{i=1}^{r} \eta_i(\theta) K_i x \quad (15)$$

where the feedback gains $K_i$, $i \in \{1,\ldots,r\}$, are determined to guarantee the following closed-loop properties:

- **Property 1:** For any admissible initial state, the states of system (12) are required to remain inside the polyhedral region described by, with $h_k \in \mathbb{R}^{n_x}$:

$$\mathcal{P}_x = \{x \in \mathbb{R}^{n_x} : \ h_k^T x \leq 1, \ k = 1,\ldots,q\}. \quad (16)$$

- **Property 2:** When $w = 0$, there exist a positive-definite function $\mathbb{V}(x) = x^T P x$, with $P > 0$ and a scalar $\tau_1 > 0$ such that $\dot{\mathbb{V}}(x) < -\tau_1 \mathbb{V}(x)$ along the trajectories of closed-loop system for any initial state in the ellipsoid $\mathcal{E}(P)$.

- **Property 3:** For $\forall w \in \mathcal{W}_\rho$ with $\rho > 0$, the closed-loop trajectories initialized in $\mathcal{E}(P)$ will always remain in this domain. Moreover, the $\mathcal{L}_\infty$ norm of the output signal $z$ satisfies: $z^T(t)z(t) \leq \gamma$, $x(0) = 0$, $t \geq 0$.

The following theorem allows for the design of a PDC controller that satisfies three above closed properties.

**Theorem 1.** Given positive scalars $\rho$ and $\tau_1$, assume there exist positive definite matrix $X \in \mathbb{R}^{n_x \times n_x}$, positive diagonal matrices $S_i \in \mathbb{R}^{n_u \times n_u}$, matrices $X_{21}^i \in \mathbb{R}^{n_z \times n_x}$, $X_{22}^i \in \mathbb{R}^{n_z \times n_z}$, $X_{23}^i \in \mathbb{R}^{n_z \times n_u}$, $X_{31}^i \in \mathbb{R}^{n_u \times n_x}$, $X_{32}^i \in \mathbb{R}^{n_u \times n_z}$, $X_{33}^i \in \mathbb{R}^{n_u \times n_u}$, $V_i \in \mathbb{R}^{n_u \times n_x}$, $W_i \in \mathbb{R}^{n_w \times n_x}$, for $i \in \{1,\ldots,r\}$ and positive scalars $\tau_2$, $\gamma$ such that the following conditions hold:

$$\begin{bmatrix} X & * \\ V_{i(l)} - W_{i(l)} & u_{\max(l)}^2 \end{bmatrix} \geq 0, \quad i, l \in \{1,\ldots,r\} \times \{1,\ldots,n_u\}, \quad (17)$$

$$\begin{bmatrix} X & X h_k \\ * & 1 \end{bmatrix} \geq 0, \quad k \in \{1,\ldots,p\}, \quad (18)$$

$$\tau_1 - \tau_2 \rho > 0, \quad (19)$$

$$\begin{bmatrix} X & * \\ C_i X & \gamma I \end{bmatrix} \geq 0, \quad i \in \{1,\ldots,r\}, \quad (20)$$

$$\begin{cases} \Psi_{ii} < 0, \quad i \in \{1,\ldots,r\} \\ \dfrac{2}{r-1}\Psi_{ii} + \Psi_{ij} + \Psi_{ij} < 0, \quad i,j \in \{1,\ldots,r\} \text{ and } i \neq j. \end{cases} \quad (21)$$

where:

$$\Psi_{ij} = \text{He}\left(\begin{bmatrix} \Psi_{ij(1,1)} & B_i^u X_{32}^j & B_i^u X_{33}^j & -B_i^u S_j & B_i^w \\ \Psi_{ij(2,1)} & -X_{22}^j & -X_{23}^j & 0 & 0 \\ V_i - X_{31}^j & -X_{32}^j & -X_{33}^j & 0 & 0 \\ W_i & 0 & 0 & -S_j & 0 \\ 0 & 0 & 0 & 0 & -\tau_2 R/2 \end{bmatrix}\right),$$

$\Psi_{ij(1,1)} = A_i X + B_i^u X_{31}^j + \tau_1 X/2$, $\Psi_{ij(2,1)} = C_i X - X_{21}^j$.

Then, controller (15) with the feedback gains given by:

$$K_i = V_i X^{-1}, \quad i \in \{1,\ldots,r\}, \quad (22)$$

guarantees the three aforementioned closed-loop properties. The proof is omitted due to the lack of space.

### B. Application to the Control of DSAS

In what follows, we provide some highlights on how to apply Theorem 1 for the control design of the DSAS. To this end, note that the system (10) has 3 scheduling variables $v_x$, $1/v_x$, $\mu(\theta_d)$ which are all measured and bounded:

$$9 \leq v_x \leq 25, \quad 0.25 \leq \mu(\theta_d) \leq 1. \quad (23)$$

By sector nonlinearity approach [15], the 8-rules T-S model (12) can be easily obtained. In this work, the control input is also bounded: $-15 \leq u \leq 15$, see Section I. The data on the disturbance (14) are $R = 1/f_{w,\max}^2$ and $\rho = 1$ where the maximal wind force is given as $f_{w,\max} = 1200N$. In addition, following the same line as in [5], both "normal driving" constraints and physical limitations on some system states should be taken into account in the control design to improve the control performance:

$$-1.75 \leq y_l + (l_f - l_s)\psi_l \leq 1.75, \quad -0.51 \leq r \leq 0.51, \quad (24)$$
$$-0.087 \leq \psi_l \leq 0.087, \quad -0.1047 \leq \dot{\delta} \leq 0.1047.$$

Since all design conditions in Theorem 1 are expressed in terms of linear matrix inequalities (LMIs). In such a way, the feedback gains $K_i$ with $i \in \{1,\ldots,8\}$ can be efficiently computed with available numerical solvers. In this work, LMI problems are done with YALMIP toolbox [16].

## V. SIMULATION VALIDATION

In this section, the new approach is validated with different driving scenarios. All numerical simulations are done using complete nonlinear vehicle model and the regulator of longitudinal speed presented in [6]. The driver model proposed in [10] is used to simulate the human driver for all test scenarios. The digital database of the "Satory" test track [6] is used to simulate the course of the simulations.

### A. Test 1: Disturbance Rejection

This scenario aims at showing the robustness performance of the designed controller. To this end, the vehicle is on a straight road section with $v_x = 15$m/s and subject to an important lateral wind force step $f_w = 1200N$ occurring from $t_1 = 70s$ to $t_2 = 76s$, see Figure 4-a. This wind force generates a yaw moment disturbance which can be felt by the driver through the steering system. It can be observed from Figure 4-b that both driving agents work in cooperation to control the vehicle. The effects of lateral wind force are effectively rejected since both lateral deviation and heading errors remain very small.

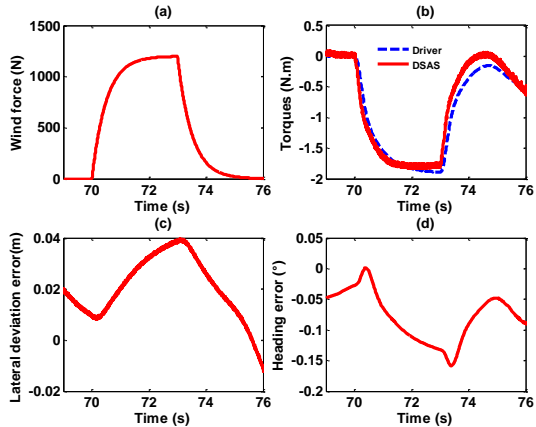

Figure 4. Shared lateral control performance in terms of disturbance rejection

### B. Test 2: Driving Task with Different Assistance Levels

This test is "designed" to show the performance of the proposed shared lateral controller under different levels of assistance according to the road curvature, see Figure 5-a. To this end, the lane tracking task is composed by four phases. In Phase 1 (from 2s to 18s), the vehicle is fully controlled by the DSAS. Figure 5-b shows that $\mu(\theta_d) = 1$ in this phase which means that maximal level of assistance is required to realize the driving task. Both controller and driver perform the lane following maneuver in Phase 2 (from 18s to 35s). During this phase, only small assistance level $\mu(\theta_d) = 0.28$ is needed. For Phase 3 (from 35s to 40s), the driver is highly involved in his/her driving task $\theta_d = 0.97$ and the E-copilot provides an important assistance $\mu(\theta_d) = 0.96$ to help him/her. This situation corresponds to the driver overload zone in Figure 1. In Phase 4 (from 40s to 55s), the assistance torque is more important than the one delivered by the driver. This driving situation corresponds to the driver underload zone in Figure 1. We can see also from Figure 6 that all state constraints considered in the control design are respected in this scenario of test.

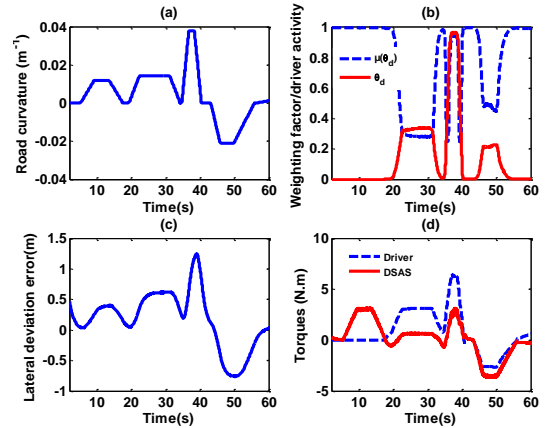

Figure 5. Driving task with different assistance levels

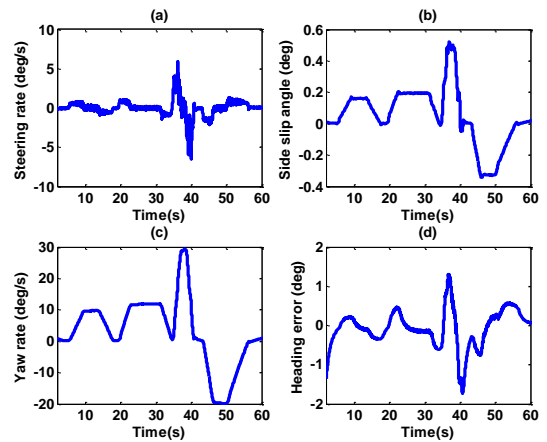

Figure 6. Control performance corresponding to Test 2

## C. Test 3: Saturation and Control Resumption of the Driver

For this test scenario, it is assumed that the vehicle is on a straight road section and the driver desires to perform a lane-change maneuver which is not specified to the controller. In the first time (from 60s to 65s), the driver provides a positive torque for the lane-change maneuver but the Driver monitoring unit indicates that the driver is in a state of distraction or tiredness ($DS=0$, see Figure 7-d). In this case, the DSAS prevents the driver to perform the maneuver with saturated control input (Figure 7-a) and maximal level of assistance (Figure 7-a). It is therefore difficult for the driver to perform this task in this state. From 65s to 75s, the driver is fully conscious of his/her driving actions ($DS=1$). We can see that although the control input is still saturated (Figure 7-a), the driver can realize the maneuver without difficulty (Figure 7-c) because of small assistance level $\mu(\theta_d)=0.25$ in this case (Figure 7-b). This test scenario points out that in many driving situations, the fictive control input can be easily saturated even if the physical limitations of the DSAS actuator are not reached (or even very small). Therefore, considering the control input in the design procedure is important for this application to achieve a good performance and also to the loss of stability.

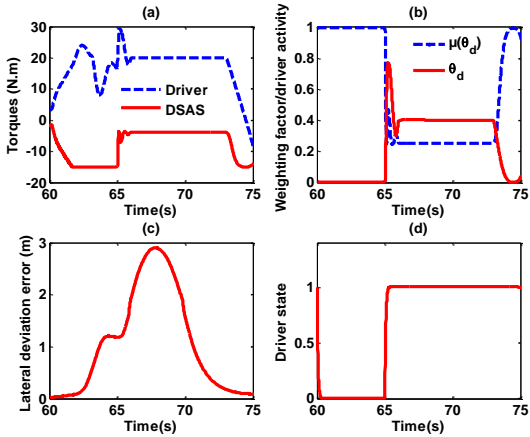

Figure 7. Input saturation and control resumption of the driver

## VI. CONCLUSIONS AND FUTURE WORKS

In this paper, a novel weighting approach has been proposed to deal with the shared lateral control problem between the driver and the DSAS. By introducing a measured weighting factor into the system, the control actions of the DSAS are computed according to the driver activity. By this way, eventual conflicts between the driver and the DSAS can be effectively managed. Based on T-S control technique, the proposed method allows for a large range of variation of the longitudinal speed. In particular, closed-loop performance under various design constraints is also guaranteed. The effectiveness of the proposed methodology is illustrated via different driving scenarios.